\documentclass[%
 aip,
 sd,%
 amsmath,amssymb,
 reprint,%
]{revtex4-1}

\usepackage{graphicx}
\usepackage{dcolumn}
\usepackage{bm}
\usepackage{epstopdf}
\usepackage{subfigure}
\usepackage{color}

\begin{document}

\preprint{AIP/123-QED}

\title[The role of cytosine methylation on charge transport through a DNA strand]{The Role of Cytosine Methylation on Charge Transport through a DNA Strand}

\author{Jianqing Qi}
\email{jqqi@uw.edu}
\affiliation{Department of Electrical Engineering, University of Washington, Seattle, Washington 98195-2500, USA}

\author{Niranjan Govind}
\email{niri.govind@pnnl.gov}
\affiliation{William R. Wiley Environmental Molecular Sciences Laboratory, Pacific Northwest National Laboratory, Richland, Washington 99352, USA}

\author{M. P. Anantram}
\email{anantmp@uw.edu}
\affiliation{Department of Electrical Engineering, University of Washington, Seattle, Washington 98195-2500, USA}

\date{\today}

\begin{abstract}

Cytosine methylation has been found to play a crucial role in various biological processes, including a number of human diseases. The detection of this small modification remains challenging. In this work, we computationally explore the possibility of detecting methylated DNA strands through direct electrical conductance measurements.  Using density functional theory and the Landauer-B$\ddot{\textup{u}}$ttiker method, we study the electronic properties and charge transport through an eight base-pair methylated DNA strand and its native counterpart. We first analyze the effect of cytosine methylation on the tight-binding parameters of two DNA strands and then model the transmission of the electrons and conductance through the strands both with and without decoherence. We find that the main difference of the tight-binding parameters between the native DNA and the methylated DNA lies in the on-site energies of (methylated) cytosine bases. The intra- and inter- strand hopping integrals between two nearest neighboring guanine base and (methylated) cytosine base also change with the addition of the methyl groups. Our calculations show that in the phase-coherent limit, the transmission of the methylated strand is close to the native strand when the energy is nearby the highest occupied molecular orbital level and larger than the native strand by 5 times in the bandgap. The trend in transmission also holds in the presence of the decoherence with the same rate. The lower conductance for the methylated strand in the experiment is suggested to be caused by the more stable structure due to the introduction of the methyl groups. We also study the role of the exchange-correlation functional and the effect of contact coupling by choosing coupling strengths ranging from weak to strong coupling limit. 

\end{abstract}

\pacs{87.14.gk, 73.63.-b, 87.15.A-}
\keywords{Charge transport, Cytosine methylation, Density functional theory, Green's function}
                              
\maketitle

\section{\label{introduction}Introduction}

DNA is best known for its essential role in biology as it carries the genetic information of living beings \cite{Endres2004}. During the last few decades, the possibility of using DNA in nanotechnology has been gaining attention from chemists, physicists, as well as engineers. The electronic properties of DNA can lead to applications in many areas. Researchers have shown that DNA has the potential to be used as circuit components in electrical devices such as double barrier resonant tunneling structures \cite{Adessi2003}, spin specific electron conductor \cite{Gohler2011, Xie2011}, field effect transistor \cite{Matsuo2012} and negative differential resistance device \cite{Kang2010}. In addition, the investigation of the electronic properties of DNA can bring new techniques for the third generation sequencing technologies. Current sequencing methods are mainly based on light emission \cite{Ohshiro2012}, which requires polymerase chain reaction (PCR) amplification. An alternative sequencing idea is to measure the transverse current through a single-stranded DNA as it translocates a nanogap or nanopore \cite{Ohshiro2012, Postma2010, Manrao2012, Laszlo2013, Shim2013}. The distinguishable electronic structures of the four DNA bases result in different electrical/ionic currents, which can be used in the third generation sequencing technologies. Recent experiments also reveal the possibility of detecting diseases by measuring the electrical conductance of DNA strands with methylated cytosine bases \cite{Tsutsui2011, Hihath2012, Bui2015}, which may open new perspectives in medical application.

Cytosine methylation refers to the replacement of the hydrogen atom by a methyl group at the 5th position of the pyrimidine ring \cite{Robertson2001}. This small modification has been found to be important in changing physical properties of DNA. For example, it can increase the melting temperature of the duplex \cite{Gill1974}, the molecular polarizability of the pyrimidine \cite{Sowers1987} and enhance the stability of the double strands. Recent study has also revealed that the methylation of cytosine bases has an essential effect on the separation of DNA strand \cite{Severin2011}. More importantly, it plays a crucial role in many biological and cellular processes such as gene expression regulation \cite{Doerfler1983, Vining2012}, cellular differentiation \cite{Khavari2010}, X-chromosome inactivation \cite{Goto1998}, embryonic development \cite{Li1992, Lei1996, Lister2009, Guo2014, Smith2014} and genomic imprinting \cite{Hore2007}. As a consequence, an increasing number of human diseases \cite{Robertson2005}, including cancer \cite{Das2004, Baylin2005}, have been found to be associated with the abnormal activities of DNA methylation. 

Targeting and identifying the methylated DNA is important for disease diagnosis and drug design. The main detection techniques used today involve PCR-based methods \cite{Shen2007}, which usually require chemical reactions \cite{Clark1994, Bock2010} to maintain the methylation feature. With the rapid development and progress in single-molecule electrical approach \cite{Xu2004, Chen2007, Mahapatro2007, Tsutsui2010}, new techniques utilizing the electronic properties of DNA bases have also been used to experimentally detect the methylated strands \cite{Clarke2009, Wallace2010, Manrao2011, Tsutsui2011, Hihath2012, Laszlo2013, Shim2013,  Bui2015}. These pioneering initial works show that it may be possible to detect methylated sequence/base by direct conductance measurement. 

Despite the recent progress in experiment, understanding charge transport through a methylated DNA  sequence/base from a theoretical perspective is lacking. In this work, we study charge transport through an eight base-pair methylated DNA strand and its native counterpart used in Ref. \onlinecite{Hihath2012}. We aim to provide an understanding of the effect of cytosine methylation on charge transport through DNA strand. To this end, we use the combination of density functional theory (DFT) and the Landauer-Buttiker approach. 

DFT is a powerful tool to obtain the accurate description of the electronic structure of molecules and materials. However, it is well known that in DFT calculations the results may depend on the choice of exchange-correlation (XC) functionals and basis sets. In addition, the lack of the derivative discontinuity and self-interaction errors in standard functionals like GGAs and hybrid-GGAs can lead to erroneous single-molecule transport predictions \cite{Toher2005}. In this paper, we also study the effects of methylation on the eight-base pair DNA strand with the B3LYP and CAM-B3LYP functionals. 

The remainder of the paper is organized as follows: we first present the method and model used in this work - we review the extraction of the tight-binding parameters from $ab~initio$ calculation which help formulate a physical understanding of the charge transport properties. Then, we discuss the Green's function approach, Landauer-B$\ddot{\textup{u}}$ttiker framework and the implementation of B$\ddot{\textup{u}}$ttiker probes using D'Amoto-Pastawski model. Next, we show the results from our calculations. We first look at the effect of methylation on the tight-binding parameters, including the on-site energies of the guanine and cytosine bases and hopping integrals between the first and second nearest neighboring bases. We further investigate the influence of cytosine methylation on the electronic properties of DNA by studying the transmission, conductance in both phase-coherent and decoherent cases. We also study the role of XC functionals on DNA charge transport. Then, we investigate the effect of contact coupling to understand the effects of the delocalization and self-interaction errors from XC functionals on the transmission. Finally, we summarize our findings in the last section.

\section{\label{method}Method}

Motivated by the measurement in Ref. \onlinecite{Hihath2012}, we choose the following two DNA sequences, each of which has eight base pairs: 1) the native strand, 5'-GCGCGCGC-3' (GC8), where G refers to the guanine base and C refers to the native cytosine base; 2) the methylated counterpart, 5'-GCmGCmGCmGCm-3' (GCm8), where Cm refers to the methylated cytosine base. 

Our approach involves two steps: (1) using DFT calculations to obtain the Hamiltonian and overlap matrices of DNA strands and (2) using the Landauer-B$\ddot{\textup{u}}$ttiker method and Green's function approach to compute the transmission and linear response conductance of the strands. All the DFT calculations are carried out with Gaussian 09 \cite{Frisch09} software package.  

We first generate the atomic coordinates of the B-form DNA strands with Nucleic Acid Builder \cite{Macke1998}. Because the conduction through DNA strands is mainly caused by the overlap of $\pi$ orbitals in the nitrogenous bases \cite{Eley1962} and our prior work shows that in the vicinity of the highest occupied molecular orbital (HOMO) level the transport is mainly through the bases \cite{Qi2013}, we delete the backbone to reduce the expensive computational cost and terminate the nitrogenous bases with hydrogen atoms. For the methylated strands, molecular dynamics simulations show that the methyl groups do not have large influence on the Watson-Crick base pairing for B-form DNA \cite{Temiz2012}. Thus, we assume that the methylated strand maintains the same B-form as the native one. We first replace the hydrogen atom in the 5th position with a -CH3 methyl group, then relax the methyl group at B3LYP/6-31G(d) \cite{Becke1993, Lee1988, Stephens1994, Hehre1972} level to get the optimized structure of the methylated cytosine. 

Next, we carry out DFT calculations using B3LYP XC functional and 6-31G basis set to determine the Hamiltonian and overlap matrices $H_0$ and $S_0$. After the Hamiltonian and overlap matrices are found, we apply L$\ddot{\textup{o}}$wdin transformation \cite{Lowdin1950} to the system to orthogonalize the basis,

\begin{equation}\label{Hamiltonian1}
H_1=S_0^{-\frac{1}{2}} H_0 S_0^{-\frac{1}{2}}
\end{equation}
Each diagonal sub-block of $H_1$ corresponds to the representation of a single base. We then find the eigenvalues and eigenvectors of the sub-blocks of $H_1$ and perform a unitary transformation of the Hamiltonian $H_1$ to $H$ via,

\begin{equation}\label{Hamiltonian}
H=U^{\dagger} H_{1} U	
\end{equation}
where the matrix $U$ is constructed from the eigenvectors of the diagonal sub-blocks of $H_1$. The diagonal sub-blocks of $H$ are the eigenvalues corresponding to each DNA base and off-diagonal elements represent the interaction between different localized energy levels. The representation of $H$ can be truncated to obtain the tight-binding parameters such as the on-site energies of a single DNA base and the hopping integrals between two neighboring DNA bases for the interested energy levels. Here we focus on analyzing the methylation effect on the tight-binding parameters at HOMO level and the corresponding Hamiltonian $H_{TB}$ is defined with the following equation,

\begin{widetext}
\begin{equation}\label{H_TB}
H_{TB}= \sum_{i} \epsilon_i c^{\dagger}_{i} c_i 
+ \sum_{\langle i, j \rangle} t^{intra}_{i,j} (c^{\dagger}_{i} c_j + H. c.)  
+ \sum_{\langle i, j \rangle} t^{inter}_{1i,j} (c^{\dagger}_{i} c_j + H. c.)
+\sum_{\langle \langle i, j \rangle \rangle} t^{inter}_{2i,j} (c^{\dagger}_{i} c_j + H. c.)  
\end{equation}
\end{widetext}
In the above equation, $c^{\dagger}_{i}$ ($c_i$) is the creation (annihilation) operator at the $i$th base (here, i = 1, 2, ... 16). $H.c.$ is the Hermitian conjugate. $\epsilon_i$ is the on-site energy of the $i$th base. $ t^{intra}_{i,j}$ is the intra-strand coupling between two nearest neighboring bases within one strand. $t^{inter}_{1i,j}$ is the inter-strand coupling between two nearest neighboring bases from the two complementary strands (i. e., the coupling for two bases within one base pair) and $t^{inter}_{2i,j}$ is the inter-strand coupling between two second nearest neighboring bases from the two strands. The notation $\langle i, j \rangle$ and $\langle \langle i, j \rangle \rangle$ in the summation depict the two bases in the first and second nearest position, respectively \cite{Mehrez2005}.

\begin{figure}
          \includegraphics[width=3.5in]{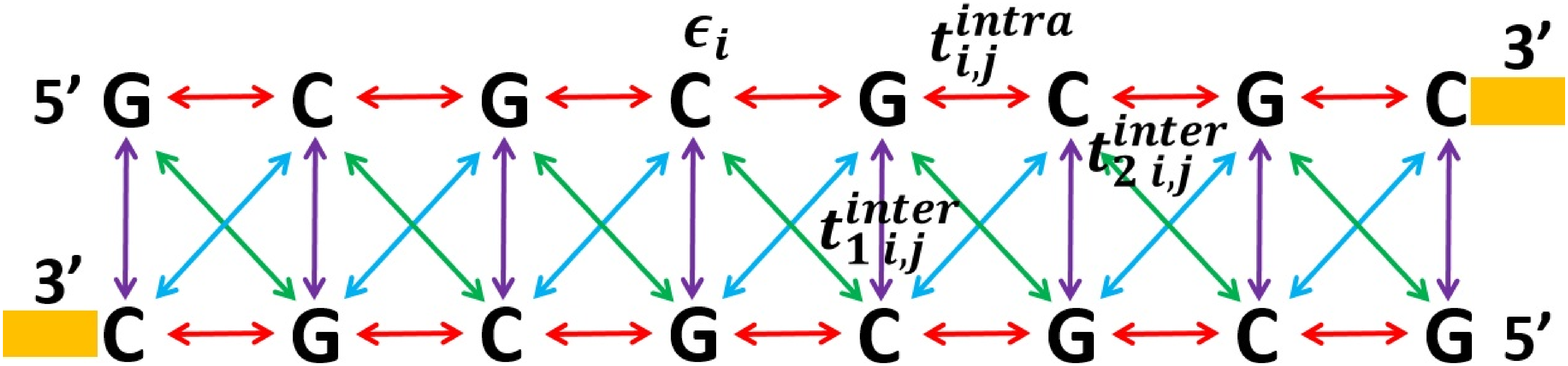}
 \caption{On-site energies and hopping integrals between two neighboring bases both within and between strands in a double-stranded DNA, with red color for the nearest intra-strand hopping, purple for the nearest inter-strand hopping, green for the second nearest inter-strand hopping from 5' end to 5' end and blue for the second nearest inter-strand hopping from 3' end to 3' end.}
    \label{Interaction}
\end{figure}

In Fig. \ref{Interaction}, we illustrate the interactions in the tight-binding model with GC8 where the contacts are shown by yellow blocks at the 3' ends as indicated. The arrows represent the hopping between two neighboring bases both within and between strands, with red color for the nearest intra-strand hopping, purple for the nearest inter-strand hopping, green for the second nearest inter-strand hopping from 5' to 5' ends and blue for the second nearest inter-strand hopping from 3' to 3' ends. The tight-binding parameters $\epsilon_i$, $ t^{intra}_{i,j}$, $t^{inter}_{1i,j}$ and $t^{inter}_{2i,j}$ have also been labeled in the corresponding position. As we will see in the later discussion, the inter-strand couplings for two second nearest neighboring bases, $t^{inter}_{2i,j}$, depends on the orientation of the strand. 

The Hamiltonian in Eq. (\ref{H_TB}) can be used to study the properties of a larger system compared to the full Hamiltonian approach since it only considers the contribution from the HOMO level. While this model can capture the essential physics of the system, to get a comprehensive understanding, the full Hamiltonian is preferable. In this work, we use full Hamiltonian shown in Eq. (\ref{Hamiltonian}) to compute the transport properties of the DNA strands, while the tight-binding Hamiltonian in  Eq. (\ref{H_TB}) is used to further our understanding of the charge transport mechanism in DNA.

The electronic transport properties of the molecule are calculated within the Landauer-B$\ddot{\textup{u}}$ttiker framework. We consider both phase-coherent and decoherent transport. In the latter case, the effect of decoherence is included using B$\ddot{\textup{u}}$ttiker probes \cite{Buttiker1986, Buttiker1988} as implemented by the D'Amoto-Pastawski model \cite{DAmato1990}. The B$\ddot{\textup{u}}$ttiker probes are fictitious probes connected to each energy level of the bases, which describes the phase-breaking of electrons due to the interaction with the environment by extracting the electron from the system and then re-injecting it back. At each B$\ddot{\textup{u}}$ttiker probe, the net current is zero. After including the B$\ddot{\textup{u}}$ttiker probes, the retarded Green's function for a DNA strand bridging across the two metal contacts is, 

\begin{equation}\label{Green}
G^r =\frac{1}{E-(H+\Sigma_{L}+\Sigma_{R}+\Sigma_{B} )}
\end{equation}
where $E$ is the energy, $H$ the Hamiltonian of the strand as shown in Eq. (\ref{Hamiltonian}), $\Sigma_{L (R)}$ and $\Sigma_{B}$ the self-energy matrices from the left (right) contact and B$\ddot{\textup{u}}$ttiker probes. 

We employ the wide-band limit (WBL) approximation to evaluate the self-energies, which ignores the real part of the matrices \cite{Herrmann2010}. Within this approximation, the self-energy due to the left (right) contact is controlled by the broadening matrix $\Gamma_{L(R)}$, and $\Sigma_{L (R)} = -i \Gamma_{L(R)}/2$.  The introduction of $\Gamma_{L(R)}$ is due to the coupling between the  DNA and the left (right) contact. The self-energy due to each fictitious B$\ddot{\textup{u}}$ttiker probe is treated in a similar manner, $\Sigma_i = -i \Gamma_i/2$, where $\Gamma_i$ is the broadening matrix arising from the coupling between the DNA and each B$\ddot{\textup{u}}$ttiker probe. The self-energy matrix $\Sigma_B$ can be obtained by summing up the self-energies from all the B$\ddot{\textup{u}}$ttiker probes, $\Sigma_B = \sum_{i=1}^{N_B}\Sigma_i$. Here $N_B$  is the number of B$\ddot{\textup{u}}$ttiker probes. In the following calculations, we treat the coupling $\Gamma_{L(R)}$ and $\Gamma_i$ to be energy-independent diagonal matrices.

With the inclusion of B$\ddot{\textup{u}}$ttiker probes, the probability for an electron to transmit from the left (L) contact to the right contact (R) is,

\begin{equation}\label{Teff}
T_{eff}=T_{LR}+\sum_{i,j=1}^{N_b}T_{L,i} W_{ij}^{-1} T_{j,R} 
\end{equation}
where $T_{ij}$ is the transmission between the $i$th site and the $j$th site in phase-coherent case, which can be determined by $T_{ij}=\Gamma_{i} G^{r} \Gamma_{j} G^{a}$, with $G^a$ being the advanced Green's function and $G^a=(G^r )^{\dagger}$. Particularly, when $ i=L$ and $j=R$, $T_{ij}$ is the phase-coherent component of the transmission between the left and right contacts. $W^{-1}$ is the inverse matrix of $W$, the elements of which are $W_{ij}=\lbrack(1-R_{ii} ) \delta_{ij}-T_{ij} (1-\delta_{ij})\rbrack$, where $R_{ii}$ is the reflection probability at probe $i$ and $R_{ii}=1 - \sum_{j \neq i}^{N}T_{ij}$. 

The small bias (linear response) conductance as a function of Fermi energy is,

\begin{equation}\label{Conductance}
G(E_f) = \frac{2e^2}{h}\int dE T_{eff}(E) \frac{-\partial f(E-E_f)}{\partial E}
\end{equation}
where $f$ is the Fermi function evaluated at 300 K.

\section{\label{results}Results and discussion}

\subsection{\label{tight_binding}Tight-binding parameters of the HOMO orbital }

\begin{figure}
 \centering
    \subfigure[]
    {
        \includegraphics[width=3.5in, height=0.92in]{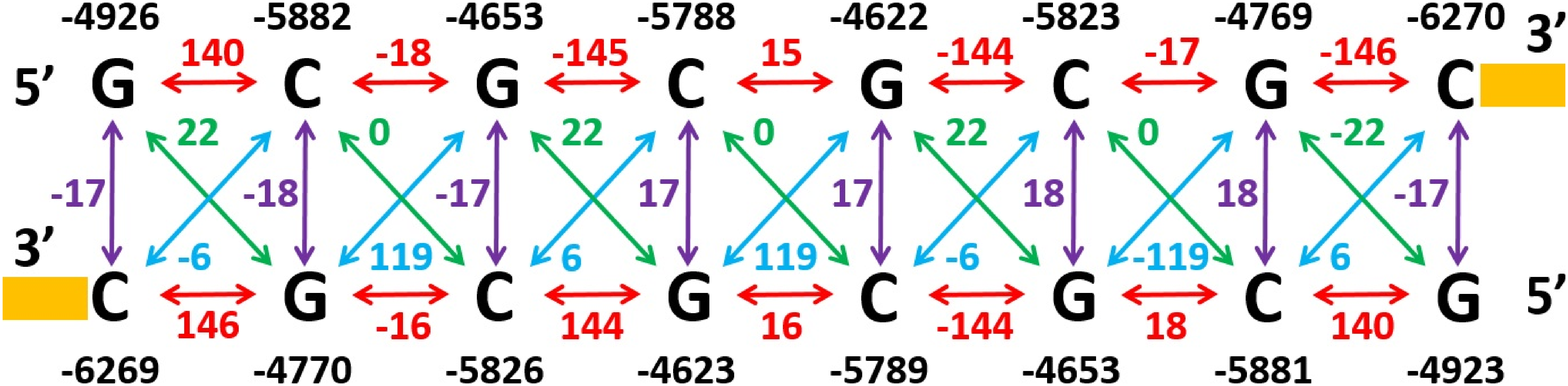}
        \label{B3LYP_TB_GC8}
    }
  \subfigure[]
    {
        \includegraphics[width=3.5in, height=0.92in]{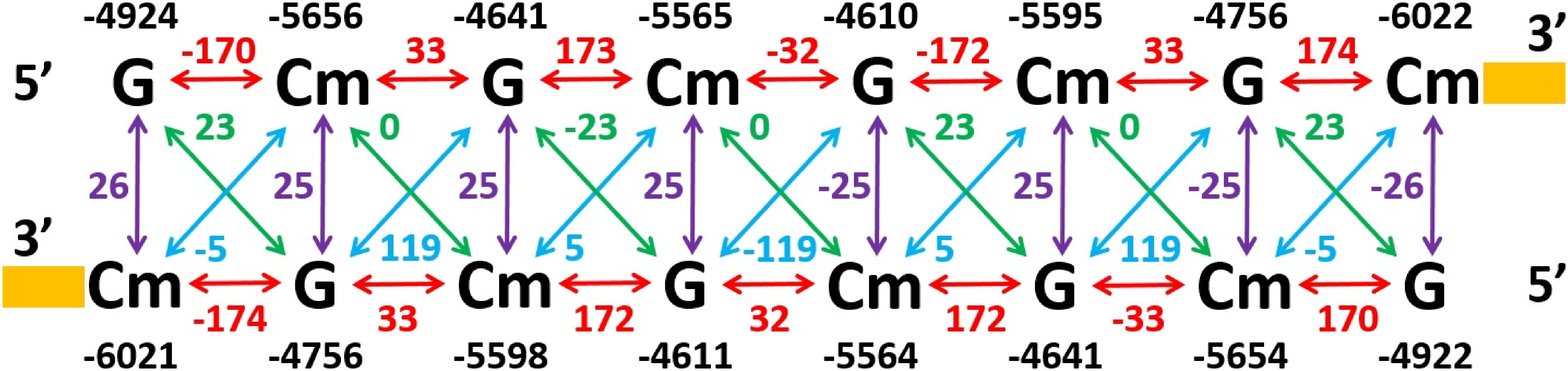}
        \label{B3LYP_TB_GCm8}
    }
    \caption{Tight-binding parameters of HOMO level obtained from DFT calculation with B3LYP/6-31G for (a) GC8 and (b) GCm8 (Unit, meV). Arrows and numbers shown in various colors follow the representations in Fig. \ref{Interaction}.}
    \label{B3LYP_TB}
\end{figure}

In this part, we analyze the tight-binding parameters of the HOMO level for GC8 and GCm8, including the on-site energy of each base and hopping integrals between two neighboring bases. On-site energies and hopping integrals are two important factors in determining the charge transport properties of a DNA strand. These parameters can be extracted from the Hamiltonian $H$ in Eq. (\ref{Hamiltonian}) using the tight-binding representation in Eq. (\ref{H_TB}). The result with B3LYP is shown in Fig. \ref{B3LYP_TB}, where the numbers shown in black are the on-site energies, the arrows and numbers shown by other different colors correspond to hopping integrals between two neighboring bases both within and between the complementary strands.

Our values of the tight-binding parameters are comparable to those in the previous work \cite{Voityuk2000, Olofsson2001, Voityuk2001, Voityuk2002, Senthilkumar2005}. The most significant differences for the parameters of GC8 and GCm8 at HOMO level lie in the on-site energies of Cs and Cms, and the hopping integrals between two nearest neighboring G and C (or Cm). For example, in Fig. \ref{B3LYP_TB} we can see that the values of the on-site energies for Cs in GC8 are around -5800 meV, while the values for Cms in GCm8 are around -5600 meV, which are approximately 200 meV higher than the native cytosine bases. The absolute values of the intra-strand hopping integrals between the two nearest neighboring G and Cm from the 5' to 3' ends and from 3' to 5' ends increase by 30 meV and 15 meV respectively, compared to GC8, while the absolute values of the inter-strand hopping integrals between the two nearest neighboring G and Cm increase by about 8 meV. It is also interesting to see that the absolute values of the intra-strand hopping between the two nearest neighboring G and C (or Cm) in the 5' to 3' and 3' to 5' directions are different by almost 9 (5) times for GC8 (GCm8), which reveals the orientation-dependent nature of electron hopping in the double helical structure.

\subsection{\label{transport}Transport properties}

The above discussions show that the small modification on cytosine bases can cause a change in tight-binding parameters at the HOMO level of a DNA strand. To understand the role of methylation in transport properties, we calculate the transmission and conductance of GC8 and GCm8 in both phase-coherent and decoherent cases, using the full Hamiltonian from the DFT calculation and focus on the energy region which is in the vicinity of the HOMO level. We set the coupling strengths to be 100 meV for $\Gamma_{L(R)}$  and 5 meV for $\Gamma_i$.

\subsubsection{\label{coherent}Phase-coherent transport}

\begin{figure}[h]
          \includegraphics[width=3.3in]{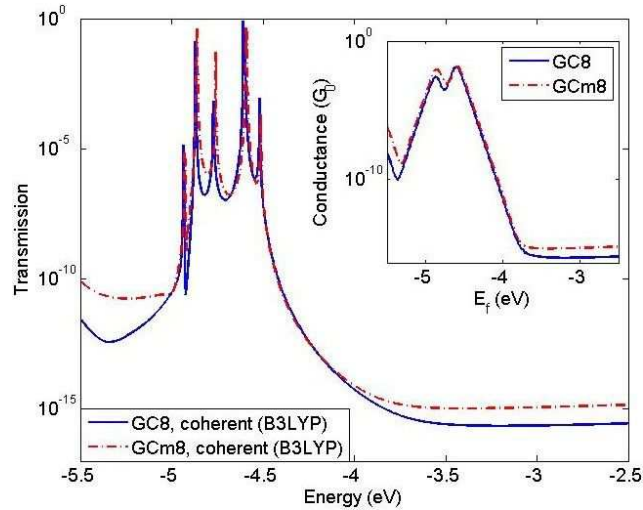}
 \caption{Transmission vs energy for GC8 (blue solid) and GCm8 (red dash) in phase-coherent transport calculated with B3LYP. Inset, conductance vs Fermi energy for GC8 (blue solid) and GCm8 (red dash).}
    \label{b3lyp_Tran_Cond_coherent}
\end{figure}

We first consider the phase-coherent transport in GC8 and GCm8. The transmission and conductance of  the two strands in the HOMO vicinity using B3LYP are shown in Fig. \ref{b3lyp_Tran_Cond_coherent} and its inset. We find that the transmission of GCm8 is very close to that of GC8 nearby the HOMO level, while away from the HOMO (when E $>$ -4 eV), the transmission of GCm8 is larger than GC8 by about 5 times in the bandgap. As one moves into the HOMO band, the transmission of both strands are oscillatory, and the transmission values for GCm8 can be larger than GC8 in some energy windows and smaller in other windows. The ratio of the transmission between GCm8 and GC8 in the energy window from -5.5 eV to -4 eV ranges from $3.37 \times 10^{-5}$ to $1.42 \times 10^{5}$ times. 

In this work we only provide a discussion based on the picture for conduction that is used in semiconductors. To proceed, we also show the linear response conductance as a function of Fermi energy in the inset of Fig. \ref{b3lyp_Tran_Cond_coherent}, with a shift in the energy axis by $|\textup{IP}_{\textup{GCm8}} - \textup{IP}_{\textup{GC8}}|$ for GCm8. We note that the conductance of GCm8 is larger than that of GC8 by about 2 times for Fermi energy locations nearby the HOMO level and 5 times away from the HOMO in the bandgap. 

It is also interesting to note that the conductance plots in the insets of both Fig. \ref{b3lyp_Tran_Cond_coherent} points to the fact that if the Fermi energy is shifted by as small as 50 meV (say due to environmental fluctuations, location of contact atoms and so on), the conductance can change by as much as 5-10 times. It is of relevance to note that in experiments the conductance fluctuates between various measurements, though the mean conductance for GC8 and GCm8 are well defined. Apart from changes in the coupling between the contact and DNA,  Fig. \ref{b3lyp_Tran_Cond_coherent} also shows that variations in the location of the Fermi level by as small as 50 meV may be responsible for the fluctuations in conductance between different measurements. A calculation that is able to account for the environmental variations which shift the Fermi energy while beyond our expertise will be useful and we expect it to yield further insight.  

\subsubsection{\label{decoherent}Phase-decoherent transport}

\begin{figure}[h]
          \includegraphics[width=3.3in]{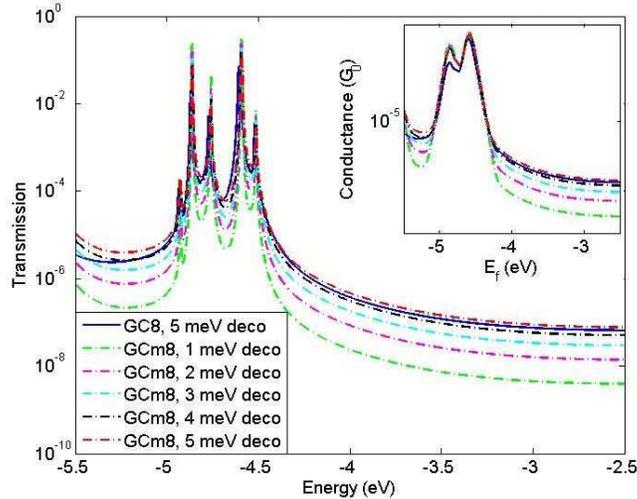}
 \caption{Transmission vs energy for GC8 and GCm8 calculated with B3LYP after including decoherence with a rate of 5 meV for GC8 and 1-5 meV for GCm8. Inset, conductance vs Fermi energy for GC8 and GCm8. Blue solid, GC8 with a decoherence rate of 5 meV; green dashed, magenta dashed, cyan dashed, black dashed and red dashed are for GCm8 with decoherence rate of 1 meV, 2 meV, 3 meV, 4 meV and 5 meV, respectively.}
    \label{b3lyp_Tran_Cond_deocherent}
\end{figure}

A few groups \cite{Gutierrez2005, Gutierrez2005_2, Gutierrez2006, Kubar2008, Zilly2010, Qi2013} have suggested that decoherence due to electron's interaction with the surrounding environment plays a large part in determining DNA conductance. We now study the influence of decoherence on charge transport through GC8 and GCm8 via the B$\ddot{\textup{u}}$ttiker probes approach to see if our conclusions in the above section change. We first set the decoherence rate to be 5 meV, a typical value in carbon nanotubes at room temperature \cite{Ishii2009, Kohler2012, Kohler2013} for both strands. The transmission and conductance for GC8 and GCm8 calculated with B3LYP after including decoherence with a rate of 5 meV is shown in blue solid and red dashed curves in Fig. \ref{b3lyp_Tran_Cond_deocherent}.

We find that for both GC8 and GCm8, after including decoherence, the peaks of the transmission become wider and the transmission values at the peaks are smaller compared to that in phase-coherent transport, due to the broadening of the localized energy levels, while away from the peaks the transmission values increase. We also note that decoherence can decrease the difference between GC8 and GCm8 for both transmission and conductance. This effect is more obvious in the HOMO-LUMO (here, LUMO refers to the lowest unoccupied molecular orbital) gap. For example, the transmission of GCm8 with 5 meV decoherence is only 1.2 times larger in the HOMO-LUMO gap, compared to 5 times in the coherent transport. Similarly, the conductance of GCm8 with the inclusion of 5 meV decoherence is 1.5 times larger than GC8 throughout the interested energy region.   

As has been pointed out by several groups, the addition of the methyl groups in the cytosine bases can increase the molecular polarizability and help stabilize the DNA structure \cite{Sowers1987, Wang1995, Sponer1997, Sugimoto2000, Derreumaux2001, Norberg2001, Rutledge2008, Fojt2009, Moser2009, Silva2010}. Using DFT calculation with B3LYP/6-31G, we find that the hydrogen bonding for a single methylated G:C base pair is 23 meV stronger than the unmethylated one. This indicates that the methylated DNA has a more rigid structure and undergoes less fluctuations in the presence of the water molecules, counter ions and other impurities, which may result in a smaller decoherence rate than the native DNA strand. To investigate the effect of the change in molecular stability, we study the transport in terms of the decoherence rate for GCm8. Since the precise value of the decohrence rate due to the structural alteration is hard to determine, we treat it as a variable. The transmission and conductance for GCm8 with decoherence rate ranging from 1 meV to 4 meV are shown in a series of dashed curves in Fig. \ref{b3lyp_Tran_Cond_deocherent}.

Interestingly, as the decoherence rate of GCm8 drops below 5 meV, the transmission of GC8 is larger than GCm8 by a few times, which is in qualitative agreement with the experimental observation  that the conductance of GC8 is 1.2 times larger than GCm8 on average. Take the energy point at -4 eV for example. As the decoherence rate of GCm8 decreases from 5 meV to 4 meV, the transmission of GC8 is now 1.2 times larger than GCm8. As the rate further decreases to 3 meV, 2 meV and 1 meV, the transmission is 2.1, 4.4, 15.7 times larger, respectively. For the conductance plots in the inset of Fig. \ref{b3lyp_Tran_Cond_deocherent}, we find the similar trend. With these analysis, we suggest that the lower conductance for GCm8 in the experimental measurement is due to the enhanced stability of the molecular structure induced by the methyl groups.

{ \color{blue} Note that the method adopted in this paper to include decoherence can qualitatively handle aspects of energy level broadening introduced by electron-vibrational coupling \cite{Kubar2008} but cannot model energy gain/loss or polaron formation \cite{Ly1999}. On the other hand, $ab~initio$ calculations of vibrational energies and the strength of their coupling to electrons is a computationally very challenging problem for large aperiodic systems such as DNA. This field has seen some welcome progress using DFT inspired tight-binding approaches \cite{Kubar2008, Gutierrez2010}, which may eventually lead to a more fundamental treatment of transport by accounting for self-consistency, DNA-contact coupling, electron-vibrational coupling and electron-environment coupling. }

\subsection{\label{result_cam_b3lyp}Transport study with CAM-B3LYP }

As DFT calculations are sensitive to the choice of the XC functionals,  it is crucial to investigate this effect on transport. In addition to B3LYP, we consider CAM-B3LYP \cite{Yanai2004} based range-separated functional. The choice of the latter is motivated by recent advances in range-separated DFT \cite{Cohen2008, Baer2010, Kronik2012, Srebro2012, Abramson2012, Lopata2013}, where the range-separation parameters are tuned to obey the Koopmans' theorem, i.e. making the HOMO as close as possible to the ionization potential (IP) of the neutral molecule. 

Rather than using the default parameters for CAM-B3LYP in Gaussian 09, we tune the relevant parameters \cite{Stein2009, Lopata2013} to satisfy Koopman's theorem, that is, to make the HOMO and IP of a neutral molecule to satisfy $\mid$HOMO + IP = 0$\mid$.

In a range-separated functional, the exchange term is described with two parts \cite{Lopata2013},

\begin{equation}\label{LC}
\frac{1}{r_{12}} = \frac{\alpha + \beta erf(\mu r_{12})}{r_{12}} + \frac{1 - [\alpha + \beta erf(\mu r_{12})]}{r_{12}}
\end{equation}
The first term is the short-range interaction, treated with DFT exchange and the second term is the long-range interaction, treated with Hartree-Fock (HF) exchange. $\alpha$ and $\beta$ are parameters which control the component of DFT and HF exchange, with $\alpha + \beta = 1$,  $0 \le \alpha \le 1$,  $0 \le \beta \le 1$ \cite{Rohrdanz2008}. We can find the optimal values of $\alpha$, $\beta$ and $\mu$ by minimizing the following function for $\mid$HOMO + IP$\mid$,

\begin{align}
J(\alpha, \mu) &= |IP_{SCF}(\alpha, \mu) - IP_{Koopmans} (\alpha, \mu)| \nonumber \\
                       &= |E^{cation}_{SCF}(\alpha, \mu) - E^{neutral}_{SCF} (\alpha, \mu) + \epsilon^{neutral}_{HOMO}(\alpha, \mu)|  \label{optimal_IP}
\end{align}

By varying the values of $\alpha$, $\beta$ and $\mu$, we find that when $\alpha = 0.2$, $\beta = 0.8$, $\mu = 0.15$, the $\mid$HOMO + IP$\mid$ function in Eq. (\ref{optimal_IP}) for GC8 can be minimized. This set of parameters is consistent with the choice of PBE-based range-separated hybrid density functional for 3,4,9,10-perylene-tetracarboxylic-dianydride (PTCDA) and 1,4,5,8-naphthalene-tetracarboxylic-dianhydride (NTCDA) molecules \cite{Abramson2012}. Since the structure of GCm8 differ from GC8 only by the methyl groups, we assume this set of parameters also works for GCm8. A comparison of the physical quantities, HOMO, IP and $\mid$IP + HOMO$\mid$ for GC8 and GCm8 calculated with functionals B3LYP and the tuned CAM-B3LYP are shown in Table~\ref{b3lyp} and Table~\ref{cam_b3lyp}, respectively.

\begin{table*}
\caption{Physical quantities computed with B3LYP/6-31G (Unit, eV).}
\label{b3lyp}
\begin{ruledtabular}
\begin{tabular}{ccccccc}

 Molecule  & HOMO  & IP   & $\mid$IP + HOMO$\mid$ \\
\hline
GC8   & -4.519   & 5.366   & 0.847\\
GCm8 & -4.512  & 5.364  & 0.852\\

\end{tabular}
\end{ruledtabular}
\end{table*}

\begin{table*}
\caption{Physical quantities computed with the tuned CAM-B3LYP/6-31G (Unit, eV).}
\label{cam_b3lyp}
\begin{ruledtabular}
\begin{tabular}{ccccccc}

 Molecule  & HOMO   & IP  & $\mid$IP + HOMO$\mid$ \\
\hline
    GC8   & -6.012   & 6.014  &   0.002  \\
    GCm8 & -6.005   & 5.990 &  0.015  \\

\end{tabular}
\end{ruledtabular}
\end{table*}

We can see that with both functionals, the HOMOs of GC8 and GCm8 are very close - only differ by 7 meV, due to the minor structural difference of the two strands. Compared to B3LYP, the effect of the tuned CAM-B3LYP is to shift the HOMO for both strands by around 1.5 eV. IP of GC8 is also similar to that of GCm8 with the two different functionals. As expected, $\mid$IP + HOMO$\mid$ for GC8 improve significantly with the tuned CAM-B3LYP functional, which indicates the efficacy of the parameters, $\alpha$, $\beta$ and $\mu$. It is also worth noting that the $\mid$IP + HOMO$\mid$ for GCm8 with the tuned CAM-B3LYP is only 15 meV, which is slightly smaller than the one calculated with B3LYP,  suggesting that the range-separation parameters determined for GC8 also hold for a different strand with a similar structure. This establishes the transferability of our work. 

\begin{figure}
 \centering
    \subfigure[ ]
    {
        \includegraphics[width=3.5in, height=0.92in]{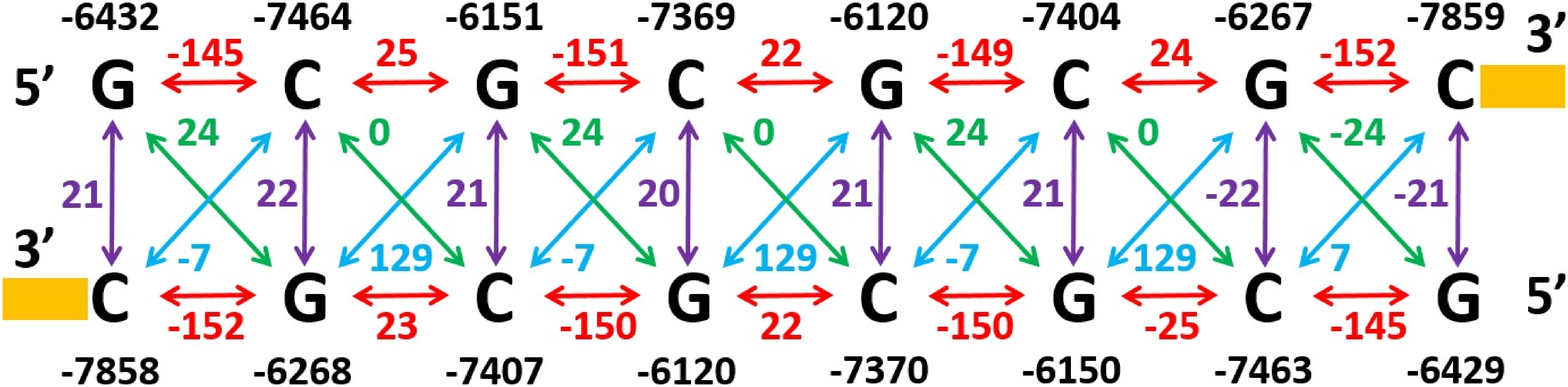}
        \label{CAM_TB_GC8}
    }
  \subfigure[ ]
    {
        \includegraphics[width=3.5in, height=0.92in]{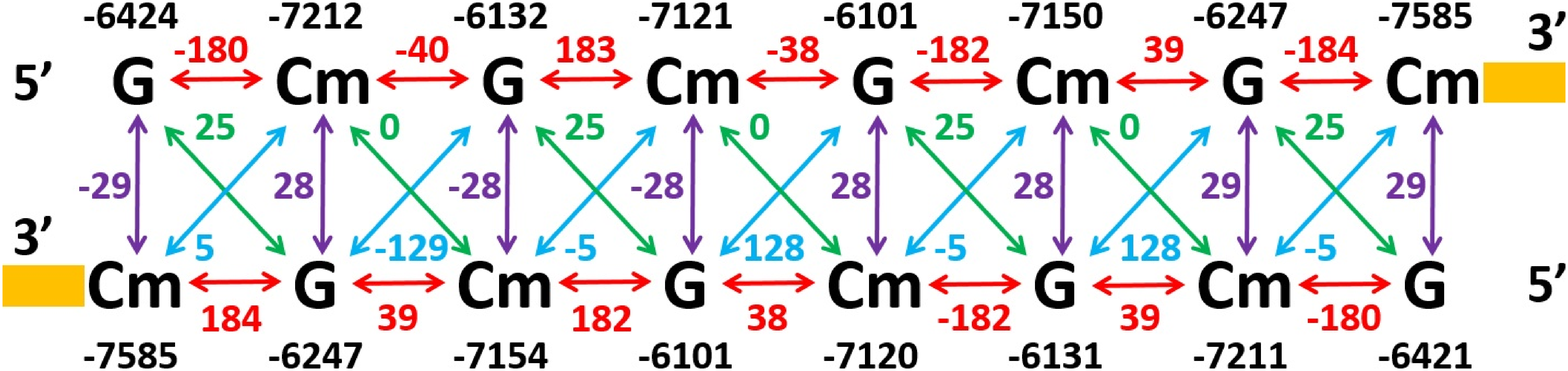}
        \label{CAM_TB_GCm8}
    }
    \caption{Tight-binding parameters of HOMO level obtained from DFT calculation with tuned CAM-B3LYP/6-31G for (a) GC8 and (b) GCm8 (Unit, meV). Arrows and numbers shown in various colors follow the representations in Fig. \ref{Interaction}.}
    \label{CAM_TB}
\end{figure}

As a comparison with the results computed with B3LYP, we also show the tight-binding parameters with tuned CAM-B3LYP in Fig. \ref{CAM_TB}. Similar to the findings with B3LYP, one can see that with the tuned CAM-B3LYP, the main influence of cytosine methylation is to change the on-site energies of cytosine bases by about 200 meV and to modify the hopping integrals between the two nearest neighboring G and C (Cm) with approximately the same amount as that in B3LYP. Comparing the quantities computed by B3LYP with the corresponding ones computed by the tuned CAM-B3LYP, we find that the main difference between the two sets of the parameters lie in the on-site energies of the G and C (Cm) bases, shifting by around 1500 meV, while the hopping integrals are of the same order of magnitude, only deviating by a few milli-electron volts.

\begin{figure}[h]
          \includegraphics[width=3.3in]{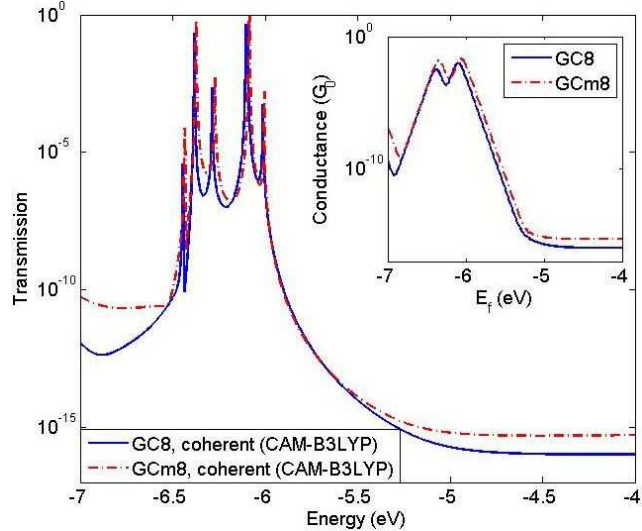}
 \caption{Transmission vs energy for GC8 (blue solid) and GCm8 (red dash) in phase-coherent transport calculated with the tuned CAM-B3LYP. Inset, conductance vs Fermi energy for GC8 (blue solid) and GCm8 (red dash).}
    \label{cam_Tran_Cond_coherent}
\end{figure}

We also show the transmission and conductance  for GC8 and GCm8 in coherent transport computed with the tuned CAM-B3LYP in Fig. \ref{cam_Tran_Cond_coherent}. The main observation in results with B3LYP also holds here with CAM-B3LYP - the transmission of GCm8 is close to GC8 nearby the HOMO level and larger than GC8 away from the HOMO. Different from B3LYP, the conductance values of GCm8 is now 8 times larger than GC8 nearby the HOMO, compared to 2 times in the above discussion for B3LYP. We attribute this finding to the larger difference in IP between GC8 and GCm8 with the tuned CAM-B3LYP, as shown in Table \ref{cam_b3lyp}.   

Comparing the transmission and conductance in Fig. \ref{b3lyp_Tran_Cond_coherent} and Fig. \ref{cam_Tran_Cond_coherent}, we conclude that the transport properties in coherent case calculated by B3LYP and the tuned CAM-B3LYP mainly differ in the energy scale. This is in agreement with the observations in the tight-binding parameters at the HOMO level shown in Fig. \ref{B3LYP_TB} and Fig. \ref{CAM_TB} that the on-site energies of DNA bases with the two different functionals differ by as much as 1500 meV, while the intra- and inter-strands hopping integrals are quite similar.

\begin{figure}[h]
          \includegraphics[width=3.3in]{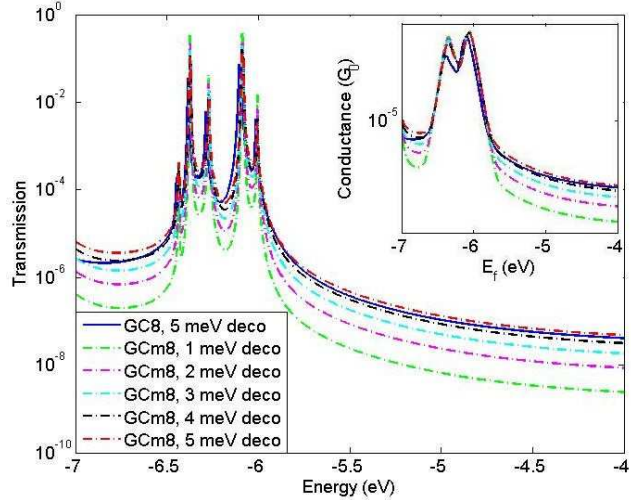}
 \caption{Transmission vs energy for GC8 and GCm8 calculated with the tuned CAM-B3LYP after including decoherence with a rate of 5 meV for GC8 and 1-5 meV for GCm8. Inset, conductance vs Fermi energy for GC8 and GCm8. Blue solid, GC8 with a decoherence rate of 5 meV; green dashed, magenta dashed, cyan dashed, black dashed and red dashed are for GCm8 with decoherence rate of 1 meV, 2 meV, 3 meV, 4 meV and 5 meV, respectively.}
    \label{cam_Tran_Cond_deocherent}
\end{figure}

In Fig. \ref{cam_Tran_Cond_deocherent}, we show the transmission and conductance for GC8 and GCm8 calculated by the tuned CAM-B3LYP with the inclusion of 5 meV decoherence for GC8 and 1-5 meV for GCm8. We find that the transmission and conductance through GC8 and GCm8 in the decoherent transport calculated with the tuned CAM-B3LYP is qualitatively similar to that with B3LYP, except the shift in the energy scale. The observation that the lower conductance for GCm8 in the experiments may be due to the more stable molecular structure of GCm8 induced by the methyl groups also hold here. It is worth pointing out that in this case the conductance of GCm8 with the decoherence rate of 5 meV is about 4 times larger than GC8 at the same energy point nearby the HOMO level, due to the fact that the IP of GCm8 is 24 meV smaller than GC8 with the tuned CAM-B3LYP. 

\subsection{\label{coupling} Effect of contact coupling}

\begin{figure}[h]
          \includegraphics[width=3.3in]{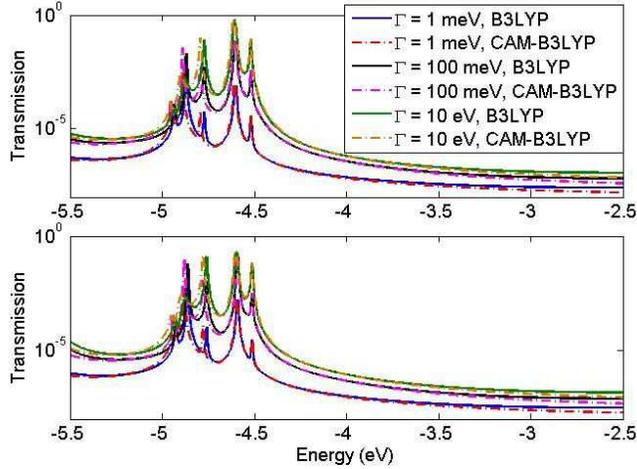}
 \caption{Transmission vs energy for GC8 (upper panel) and GCm8 (lower panel) calculated with B3LYP (solid curves) and the tuned CAM-B3LYP (dashed curves) functionals after including decoherence with a rate of 5 meV at $\Gamma = 1 meV$, $\Gamma = 100 meV$ and $\Gamma = 10 eV$. The energy scale of transmission computed with CAM-B3LYP has been shifted by $(\textnormal{HOMO}_{\textnormal{CAM-B3LYP}} - \textnormal{HOMO}_{\textnormal{B3LYP}})$.}
    \label{transport_coupling}
\end{figure}

Based on the observation in Sec. \ref{transport} and \ref{result_cam_b3lyp}, we find similar transport results with the B3LYP and the tuned CAM-B3LYP XC functionals in both phase-coherent and decoherent transports. In this section, we further study the effect of functionals on DNA charge transport with various coupling strengths to the contacts. The broadening of the molecular levels due to the coupling to the contacts plays an important role in determining the electron flow through a molecule \cite{Zahid2003}. According to the tight-binding parameters in Sec. \ref{tight_binding} and \ref{result_cam_b3lyp}, the hopping integral value in a DNA molecule ranges from a few milli-electron volts to hundreds of milli-electron volts. A normal coupling strength due to the electrical contacts should be comparable to the hopping integral. Thus, we choose three values of coupling strengths, $\Gamma = $ 1 meV (weak coupling limit), 100 meV (normal coupling) and 10 eV (strong coupling limit) and assume $\Gamma_L = \Gamma_R = \Gamma$ to study the transmission of GC8 and GCm8 in decoherent transport with a decoherence rate of 5 meV, as shown in Fig. \ref{transport_coupling}.    

To quantify our understanding, the energy scale for transmission computed with the tuned CAM-B3LYP has been shifted by $(\textnormal{HOMO}_{\textnormal{CAM-B3LYP}} - \textnormal{HOMO}_{\textnormal{B3LYP}})$ in Fig. \ref{transport_coupling}. For GC8, as the coupling strength increases, the transmission increases drastically with both functionals. This effect is more evident when the coupling strengths are in weak regime. For example, as the coupling strengths increases from 1 meV to 100 meV, the transmission of GC8 is enhanced by as much as 25 times. 

It is reasonable to assume that, in the weak molecule-lead coupling limit, the misaligned molecular levels (for example, from a B3LYP calculation) with respect to the leads would result in an incorrect molecular response as charge is transferred from the leads into the molecule which should be observable. However, our transmission results with both B3LYP and the tuned CAM-B3LYP yield almost identical results in the vicinity of the HOMO irrespective of the strength of the contact coupling even though the delocalization and self-interaction errors are minimized in tuned range-separated XC forms \cite{Baer2010, Kronik2012, Srebro2012, Abramson2012, Lopata2013}. We find similar results for GCm8. The natural follow-on question is how large is this effect?, and if it is observable in realistic systems. The other factor could be the inherent approximations within the Green's function-DFT approach itself. In order to quantify these effects, a systematic study of charge transport through well-characterized molecular junctions with different XC potentials and range-separation in the weak and strong coupling limits is needed.

\section{\label{conclusion}Conclusion}

In summary, we have studied the possibility of detecting a methylated DNA strand using its electronic properties from a theoretical perspective, based on DFT calculations and the Landauer-B$\ddot{\textup{u}}$ttiker approach. We explore the electronic structures and charge transport properties through an eight base-pair methylated DNA strand and its native counterpart used in the experiment of Ref. \onlinecite{Hihath2012}. We first look at the tight-binding parameters of the HOMO level for the two strands and find that the methylation has a large influence on the on-site energies of cytosine bases and hopping integrals between two nearest neighboring guanine and cytosine bases. The on-site energies of the methylated cytosine bases are approximately 200 meV higher than that of native cytosine bases. The intra-strand hopping integrals between two nearest neighboring guanine and cytosine bases from 5' to 3' ends for a methylated strand are about 30 meV stronger than that of native strand, and 15 meV stronger from 3' to 5' ends. We then carry out transport calculations in both phase-coherent and decoherent cases. Our calculations show that in phase-coherent transport, the transmission of electrons close to the HOMO band of the two strands are very similar. At energies away from HOMO, the transmission of GCm8 is larger than GC8. The linear response conductances of GC8 and GCm8 are also close to each other, with the conductance of GCm8 being slightly higher than GC8. The calculations also show that the conductance of these molecules can change by 5-10 times if there is a small shift (50 meV) in the Fermi energy due to environmental variations from experiment to experiment. In the phase-decoherent case, we find that the transmission and conductance trends occuring in the phase-coherent limit also hold in the HOMO vicinity if the decoherence rates for the two strands are the same. However, if the decoherence rate of the methylated strand is lower, we observe that the transmission and conductance of the methylated strand is also lower than the native strand, which is consistent with the experimental findings. Our results show that the effect of the two different functionals is to alter the on-site energies of the DNA bases at the HOMO level, while the transport properties do not depend much on the two functionals.

\begin{acknowledgments}
The authors thank Professor Joshua Hihath at University of California, Davis for extensive discussions on experimental conductance measurements of DNA strands. The authors also thank Benjamin E. Van Kuiken and Professor Xiaosong Li at University of Washington, Dr. Fernando R. Clemente at Gaussian Inc. for their help with CAM-B3LYP. Niranjan Govind and M. P. Anantram would like to thank the Northwest Institute for Advanced Computing (NIAC) workshop at PNNL (NIAC Day@PNNL in March 2014) for stimulating discussions that lead to this collaboration. M. P. Anantram and Jianqing Qi acknowledge the support from the National Science Foundation under Grant No. 102781. 
\end{acknowledgments}


\end{document}